\documentclass[a4paper]{article}
\usepackage{Odyssey2020}
\usepackage{epsfig,amssymb,amsmath}
\usepackage{graphicx, epstopdf, dblfloatfix, subfig}
\usepackage{balance, url}
\usepackage{multirow, multicol}
\usepackage{booktabs}
\usepackage[bottom]{footmisc}
\ninept

\title{The 2021 NIST Speaker Recognition Evaluation}

\makeatletter
\def\name#1{\gdef\@name{#1\\}}
\makeatother
\name{{\em Seyed Omid Sadjadi$^{\textrm{1}}$, Craig Greenberg$^\textrm{1}$, Elliot Singer$^{\textrm{2},\dagger}$,}\\
	{\em Lisa Mason$^\textrm{3}$, Douglas Reynolds$^\textrm{3}$} \thanks{$^\dagger$The work of MIT Lincoln Laboratory is supported by the Department of Defense under Air Force Contract FA8702-15-D-0001. Any opinions, findings, conclusions or recommendations expressed in this document are those of the authors do not necessarily reflect the views of the Department of Defense.}}
\address{
	$^\textrm{1}$NIST ITL/IAD/Multimodal Information Group, MD, USA\\
	$^\textrm{2}$MIT Lincoln Laboratory, MA, USA\\
	$^\textrm{3}$U.S. Department of Defense, MD, USA \\
	{\small \tt craig.greenberg@nist.gov} }

\graphicspath{{./figs/}}

\begin{document}
	\maketitle

	\begin{abstract}
		The 2021 Speaker Recognition Evaluation (SRE21) was the latest cycle of the ongoing evaluation series conducted by the U.S. National Institute of Standards and Technology (NIST) since 1996. It was the second large-scale multimodal speaker/person recognition evaluation organized by NIST (the first one being SRE19). Similar to SRE19, it featured two core evaluation tracks, namely audio and audio-visual, as well as an optional visual track. In addition to offering \textit{fixed} and \textit{open} training conditions, it also introduced new challenges for the community, thanks to a new multimodal (i.e., audio, video, and selfie images) and multilingual (i.e., with multilingual speakers) corpus, termed WeCanTalk, collected outside North America by the Linguistic Data Consortium (LDC). These challenges included: 1) trials (target and non-target) with enrollment and test segments originating from different domains (i.e., telephony versus video), and 2) trials (target and non-target) with enrollment and test segments spoken in different languages (i.e., cross-lingual trials). This paper presents an overview of SRE21 including the tasks, performance metric, data, evaluation protocol, results and system performance analyses. A total of 23 organizations (forming 15 teams) from academia and industry participated in SRE21 and submitted 158 valid system outputs. Evaluation results indicate: audio-visual fusion produce substantial gains in performance over audio-only or visual-only systems; top performing speaker and face recognition systems exhibited comparable performance under the matched domain conditions present in this evaluation; and, the use of complex neural network architectures (e.g., ResNet) along with angular losses with margin, data augmentation, as well as long duration fine-tuning contributed to notable performance improvements for the audio-only speaker recognition task.
	\end{abstract}

	\section{Introduction}
	
	The United States National Institute of Standards and Technology (NIST) organized the 2021 Speaker Recognition Evaluation (SRE21) in the summer--fall of 2021. It was the latest cycle in the ongoing series of speaker recognition technology evaluations conducted by NIST since 1996 \cite{nistsre,twodecades}. The objectives of the evaluation series are 1) for NIST to effectively measure system-calibrated performance of the current state of technology, 2) to provide a common test bed that enables the research community to explore promising new ideas in speaker recognition, and 3) to support the community in the development of advanced technology incorporating these ideas.
	
	Following the audio-visual SRE19~\cite{nistsre19av}, SRE21 was the second large-scale multimodal (i.e., audio-visual) speaker/person recognition evaluation organized by NIST. Similar to SRE19, it featured two core evaluation tracks, namely audio and audio-visual, as well as an optional visual-only track. Table~\ref{tbl:tracks} summarizes the tracks for SRE21. The audio-visual material used in SRE21 was extracted from a new multimodal and multilingual corpus, titled WeCanTalk~\cite{wecantalk}, collected outside North America by the Linguistic Data Consortium (LDC). The WeCanTalk corpus consists of phone calls, video recordings, and closeup images (e.g., selfies) collected from multilingual subjects. This paper presents an overview of SRE21 including the tasks, the performance metric, data, and the evaluation protocol as well as results and performance analyses of submissions. As in SRE19, the NIST conversational telephone speech (CTS) speaker recognition challenge served as a prerequisite for SRE21, meaning that in order for one to participate, they must have first completed the challenge (i.e., submitted at least one valid system output to the CTS Challenge platform\footnote{\url{https://sre.nist.gov/cts-challenge}}). SRE21 was coordinated entirely online using the NIST SRE web platform\footnote{\url{https://sre.nist.gov/sre21}} that supports a variety of evaluation related services such as registration, data license agreement management, data distribution, system output submission and validation/scoring, and system description uploads.
	
	\begin{table*}[h]
	\caption{\it SRE21 tracks}
	\vspace{-6mm}
    \begin{center}
    \begin{tabular}{|l|c|c|c|}
     \hline
    \textbf{Track} & \textbf{Enrollment} & \textbf{Test}  & \textbf{Submission required} \\ 
     \hline
     \hline
    Audio & CTS/AfV & CTS/AfV & Yes \\ 
      \hline
    Visual & Close-up Image & Video & No \\
      \hline
    Audio-Visual & CTS + Close-up Image & Video & Yes \\ 
    \hline
    \end{tabular}
    \vspace{-4mm}
    \label{tbl:tracks}
    \end{center}
    \end{table*}
	
	Although SRE21 was organized in a similar manner to SRE19, it introduced several new features. First, in addition to audio from video (AfV) data (first introduced in SRE18~\cite{nistsre18_is19}), SRE21 also included CTS data. Second, thanks to the multimodal and multilingual data in the WeCanTalk corpus, it involved cross-lingual and cross-domain target and non-target trials. In the cross-lingual scenario, the enrollment and test segments were spoken in different languages, while in the cross-domain (or cross-source) scenario the enrollment and test segments originated from different source types (i.e., CTS and AfV). Third, in the visual-only and audio-visual tracks, closeup images (selfies) of the target individuals were provided for face enrollment. Fourth, unlike SRE19 which used full video recordings with speech durations ranging from 10 seconds to 600 seconds, SRE21 introduced speech duration variability in the test segments (in the [10s, 60s] range) and limited the speech duration for enrollment to approximately 60 seconds. Additionally, SRE21 involved segments with free-style (mixed-language) conversational speech. Finally, unlike SRE19 which only offered an \textit{open} training condition, SRE21 offered both \textit{open} and \textit{fixed} training conditions. The \textit{fixed} training condition, which designates a common set of data for system training and development purposes, allows for uniform and meaningful algorithmic comparison of systems. On the other hand, the \textit{open} training condition, which allows the use of any publicly available and/or proprietary data for system training and development, was offered to demonstrate possible performance gains that could be achieved with unconstrained amounts of data. System submission was required for the \textit{fixed} condition, and optional, yet highly encouraged, for the \textit{open} condition. The NIST designated data for the \textit{fixed} condition included the SRE16~\cite{nistsre16_is17} evaluation set, the JANUS Multimedia Dataset~\cite{sell2018}, an SRE21 in-domain development set with audio-visual data from 20 individuals, as well as a newly developed large-scale telephony dataset called the NIST SRE CTS Superset~\cite{sadjadi2021nist} with more than 600,000 segments from 6867 speakers.
	 
\begin{figure}[!t]
		\centering
		\includegraphics[width=\linewidth, clip, trim=3mm 0mm 0mm 0mm]{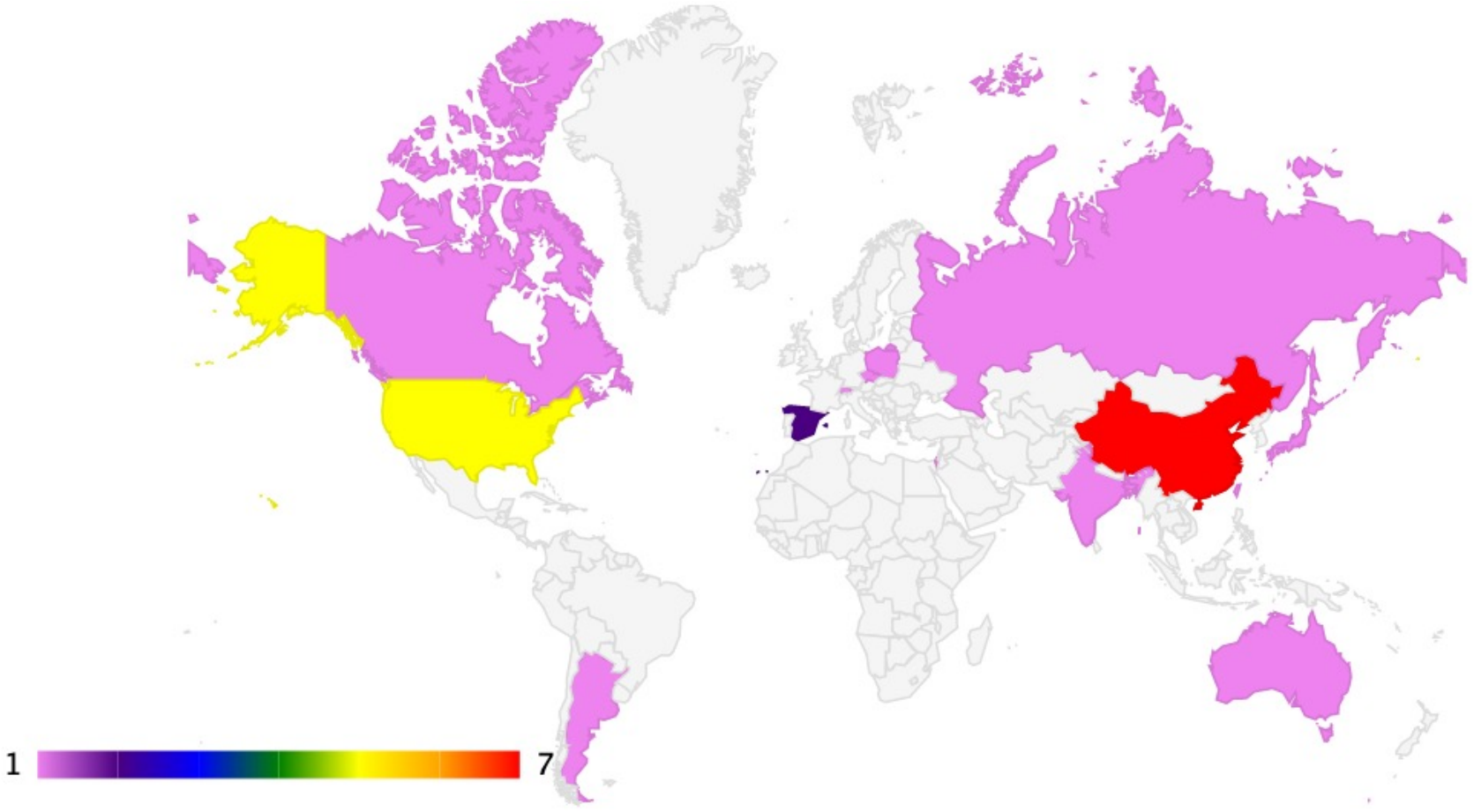}
		\vspace{-4mm}
		\caption{{\it Geographic heatmap for SRE21 participating sites. Colors denote the number of sites per country.}}
		\label{fig:map}
		\vspace{-4mm}
	\end{figure}

\begin{figure}[b]
		\centering
		\includegraphics[width=\linewidth, clip, trim=0mm 30mm 0mm 0mm]{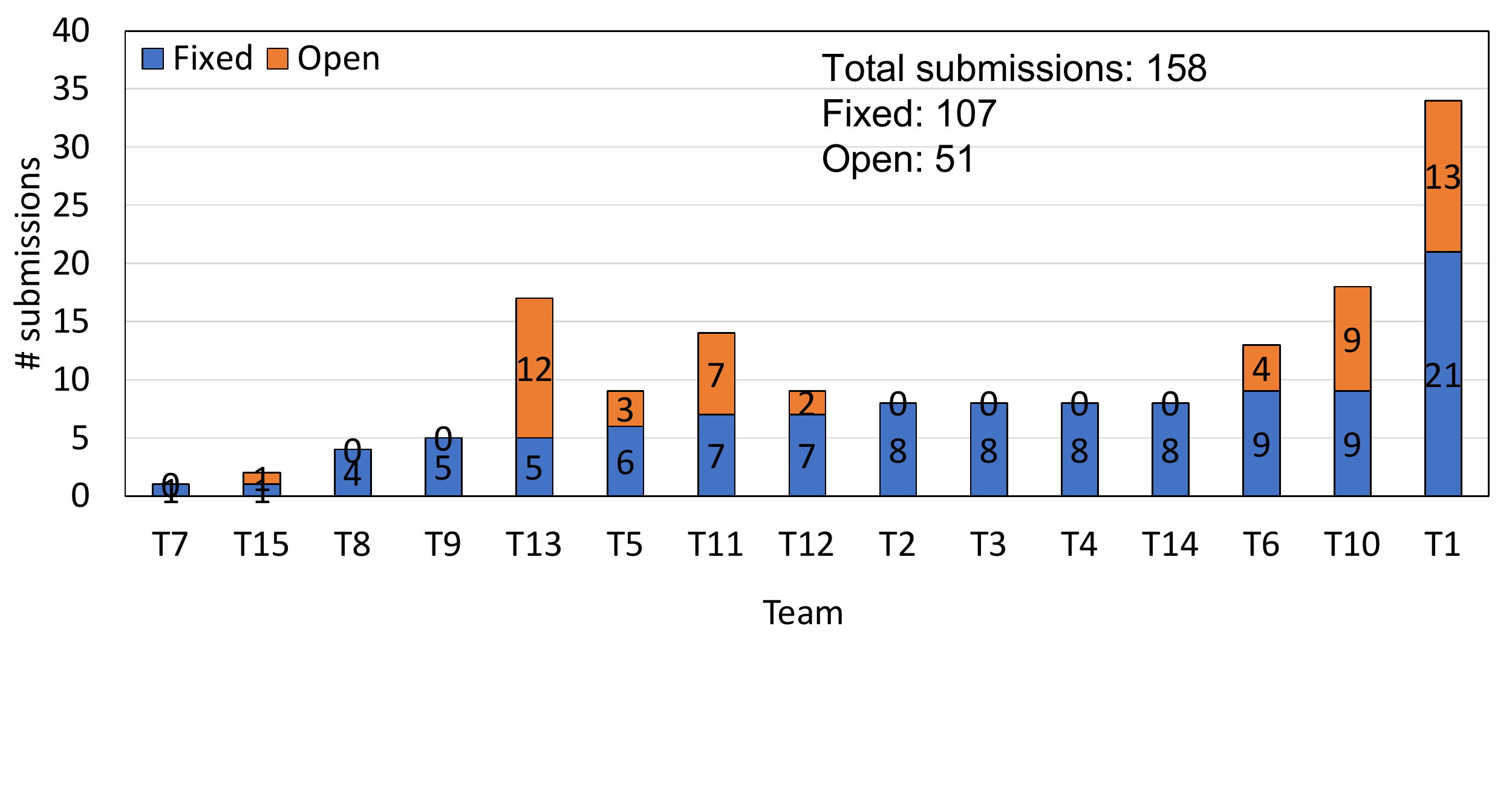}
		\vspace{-6mm}
		\caption{{\it SRE21 submission statistics.}}
		\label{fig:submission_stats}
	\end{figure}

	 To participate in SRE21, teams could register up to three systems for each track (i.e., audio, audio-visual, and visual), one of which under each track should have been designated as the primary system, and the other two as either contrastive or single best systems. Teams could make a total of 9 submissions for each of the three systems until the evaluation period was over, with up to 3 submissions per day. Over the course of the evaluation, which ran from August through October, 2021, a total of 15 teams, 10 of which were led by industrial institutions, from 23 sites made 158 valid submissions (note that the participants processed the data locally and submitted only the output of their systems to NIST for uniform scoring and analysis purposes). Figure~\ref{fig:map} displays a geographic heatmap representing the number of participating sites per country. All participant information, including country, was self-reported. The overall number of submissions in SRE21, and a breakdown for the \textit{fixed} condition, per team per track (i.e., audio, visual, and audio-visual) is shown in Figure~\ref{fig:submission_stats}.
	
	As in the most recent SREs, and in an effort to provide reproducible state-of-the-art baselines for SRE21, NIST released a report \cite{sre21_baseline} containing descriptions of speaker and face recognition baseline systems as well as results obtained using these standalone state-of-the-art (as of SRE19) deep neural network (DNN) embedding based systems as well as their fusion (see Section~\ref{sec:baseline} for more details). 
	
	\section{Task Description}
	
	The primary task for SRE21 was \textit{speaker/person detection}, meaning that given a test audio or video segment and a target individual's enrollment audio or selfie image, automatically determine whether the target person is present in the test segment. The test segment along with the enrollment data from a designated target individual constitute a \textit{trial}. The system is required to process each trial independently and to output a log-likelihood ratio (LLR), using natural (base $e$) logarithm, for that trial. The LLR for a given trial including a test segment $s$ is defined as follows

    \begin{equation}
    LLR(s) = \log \left(\frac{P\left(s|H_0\right)}{P\left(s|H_1\right)} \right).
    \end{equation}
    where $P\left(\cdot\right)$ denotes the probability density function (pdf), and $H_0$ and $H_1$ represent the null (i.e., the target individual is present in $s$) and alternative (i.e., the target individual is not present in $s$) hypotheses, respectively.
	
	\setlength{\tabcolsep}{2.2mm}
	\renewcommand{\arraystretch}{1.1}
	\begin{table*}[t]
		\caption{\it Data statistics for SRE21 development (DEV) and TEST subsets per track. M and F denote male and female, respectively.}
		\label{tab:data_stats}
		\centering
		\begin{tabular}{llccccc}
			\toprule
			\textbf{Track} & \textbf{Subset} & \textbf{\#speakers (M / F)} & \textbf{\#Enroll segments} & \textbf{\#Test segments} & \textbf{\#Target trials} & \textbf{\#Non-target trials} \\
			\midrule
			\multirow{2}{*}{Audio} & DEV & 5 / 15 & 138 & 2001 & 14,458 & 177,793 \\
			& TEST & 43 / 139 & 1247 & 17,037 & 132,038 & 5,899,731 \\
			\hline
			\multirow{2}{*}{Visual} & DEV & 5 / 15 & 20 & 388 & 388 & 4572 \\
			& TEST & 43 / 139 & 182 & 3177 & 3166 & 279,844 \\
			\hline
			\multirow{2}{*}{Audio-Visual} & DEV & 5 / 15 & 138 & 388 & 2696 & 36,208 \\
			& TEST & 40 / 136 & 1247 & 3177 & 22,999 & 1,073,892\\
			\bottomrule
		\end{tabular}
		\vspace{-4mm}
	\end{table*}
	
	\section{Data}
	\label{sec:data}
	
	
	\subsection{Training set}
	As noted previously, unlike in SRE19 that only offered an \textit{open} training condition, SRE21 offered both \textit{fixed} and \textit{open} training conditions. The re-introduction of the \textit{fixed} training condition was based on the feedback received during the discussion sessions at the SRE19 post-evaluation workshop to facilitate meaningful and fair cross-system comparisons in terms of core speaker recognition algorithms/approaches (as opposed to particular data) used.
	
	For the \textit{fixed} training condition, in which participation was required, NIST designated a \textit{common} set for system training and development purposes consisting of the following corpora:
	
  \begin{itemize}
  \item NIST SRE CTS Superset~\cite{sadjadi2021nist} (LDC2021E08~\cite{supersetldc})
  \item 2016 NIST SRE Evaluation Set (LDC2019S20~\cite{sre16ldc})
  \item 2021 NIST SRE Development Set (LDC2021E09~\cite{sre21ldc})
  \item JANUS Multimedia Dataset~\cite{sell2018} (LDC2019E55~\cite{janusldc})
  \end{itemize}
    Except for the CTS Superset which was available directly from the SRE web platform (\url{https://sre.nist.gov}), all datasets were available from the LDC subject to the approval of the data license agreement. In addition to these data, participants could also use the VoxCeleb corpus\footnote{\url{http://www.robots.ox.ac.uk/~vgg/data/voxceleb/}}. Publicly available, non-speech audio and data, e.g., noise samples, room impulse responses (RIR), and filters, could also be used, provided that a clear description was given in the final system description report. The use of pretrained speech models on data other than what was designated above was not allowed for the \textit{fixed} condition; however, participating teams could use pretrained models for the visual tracks (or sub-tracks), e.g., for face detection and face encoding extraction.

    For the \textit{open} training condition, in which participation was optional but highly encouraged, teams were allowed to utilize additional proprietary and/or publicly
    available data for system training and development. Participating teams were required to provide a sufficient description of audio (speech and non-speech) and visual data resources as well as pre-trained models used during the training and development of their systems in the \textit{open} condition.

    \subsection{Development and test sets}
	
	The SRE21 development (DEV) and test (TEST) sets were extracted from a new multimodal and multilingual corpus, titled WeCanTalk~\cite{wecantalk}, collected by the LDC. The WeCanTalk corpus is composed of phone calls and video recordings collected outside North America, spoken in Cantonese, English, and Mandarin. Recruited subjects (aka \textit{claques}) for the collection made multiple calls to people in their social network (e.g., family, friends), and recorded videos of themselves while performing various activities (e.g., amateur Vlog style videos). They also supplied close-up images of their faces (e.g., selfies). The CTS segments extracted from WeCanTalk were encoded as a-law sampled at 8~kHz in SPHERE formatted files, while the AfV segments were encoded as 16-bit FLAC files sampled at 16~kHz. All video data were encoded as MPEG4.
	
	Compared to the video data in the VAST corpus~\cite{vast} that was used for the audio-visual SRE19, the video data in the WeCanTalk exhibits less variations in terms of acoustic and visual conditions, and the majority of the videos contain monologues. Therefore, in order to make the conditions and tracks associated with the video data more challenging, we introduced test segment duration variability for both the AfV and video data in the audio, visual, audio-visual tracks. More specifically, the AfV enrollment segment speech duration was limited to 60 seconds, and the AfV test segment speech duration was sampled from the uniform distribution [10~s, 60~s]. This is the same approach that has been applied to the CTS data since SRE16~\cite{nistsre16_is17} for creating the enrollment and test segments. 
    
	
	To assist teams in their system development efforts, NIST released a small in-domain DEV set that mirrored the evaluation conditions. The DEV set included audio and video segments as well as close-up images from 20 individuals randomly selected from WeCanTalk. There was no overlap between the DEV and TEST sets in SRE21. Table~\ref{tab:data_stats} summarizes the data statistics for the SRE21 DEV and TEST subsets per track.
	
	In terms of the enrollment condition, which is defined as the number of speech segments or images provided to create a target speaker/person model, SRE21 involved two scenarios: namely 1-segment and 3-segment enrollment. In the 1-segment scenario, the system was given only one audio segment or image to build the model of the target speaker/person. For audio (i.e., CTS and AfV) trials, one segment containing approximately 60 seconds\footnote{As determined by a speech activity detector (SAD) output.} of speech was provided, while for visual trials, a close-up image of the target individual (e.g., a selfie) was provided. For the 3-segment condition, the system was given three segments, each containing approximately 60 seconds of speech, to build the model of the target speaker, all from the same phone number. This condition only involved the CTS data. Note that trials involving this condition were excluded from the official metric for SRE21 for increased challenge.
	
	SRE21 offered the following test conditions:

    \begin{itemize}
    \item Trials involving CTS data were conducted with test segments from both same and different phone numbers as the enrollment segment(s).
    \item Trials (target and non-target) involving audio data were conducted with test segments spoken both in the same and different languages as the enrollment segment(s).
    \item Trials (target and non-target) involving audio data were conducted with test segments originating from both same and different source type (i.e., CTS vs AfV) as the enrollment segment(s).
    \item There were no cross-gender trials.
    \item Each test video contained audio-visual data from only a single individual.
    \end{itemize}

    For AfV trials in the audio-only track, NIST extracted and released audio segments from videos; however, participants were responsible for extracting the relevant audio-visual data (i.e., speech or face frames) from videos for the visual-only and audio-visual tracks.

    As in the most recent evaluations, gender labels were not provided for the enrollment/test segments in the TEST set.

	\section{Performance Measurement}
	\label{sec:metric}
	
	The primary performance measure for SRE21 was a detection cost defined as a weighted sum of false-reject (miss) and false-accept (false-alarm) error probabilities. Equation (\ref{eq: cdet}) specifies the SRE21 primary normalized cost function for some decision threshold $\theta$,
	\vspace{-4mm}
	
	\begin{equation} \label{eq: cdet}
	C_{norm}\left(\theta\right) = P_{miss}\left(\theta\right) + \beta \times P_{fa}\left(\theta\right) ,
	\end{equation}
	where $\beta$ is defined as
	\begin{equation}
	\beta = \frac{C_{fa}}{C_{miss}} \times \frac{1-P_{target}}{P_{target}}.
	\end{equation}
	The parameters $C_{miss}$ and $C_{fa}$ are the cost of a missed detection and cost of a false-alarm, respectively, and $P_{target}$ is the \textit{a priori} probability that the test segment speaker is the specified target speaker. The primary cost metric, actual $C_{primary}$ for SRE21 was the average of normalized costs calculated at two operating points along the detection error trade-off (DET) curve \cite{nist1997}, with $C_{miss}=C_{fa}=1$, $P_{target_1}=0.01$ and $P_{target_2}=0.05$. Here, $\log(\beta)$ was applied as the detection threshold $\theta$ where log denotes the natural logarithm. Additional details can be found in the SRE21 evaluation plan \cite{sre21evalplan}.
    
    Based on the available metadata, the trials in the audio track were divided into a number of partitions. Each partition is defined as a combination of: speaker/person gender (male vs female), data source match (Y vs N), language match (Y vs N), enrollment/test phone number match (Y vs N).

    As for the audio-visual track, trials were partitioned based on the metadata available for the audio component (i.e., gender and language match). The visual trials were only equalized based on gender. 
    
    Accordingly, the actual $C_{Primary}$ was calculated for each partition, and the final result was the average of all $C_{Primary}$'s across the various partitions. More information about the partitions in SRE21 Audio TEST set can be found in Table~\ref{tab:partitions}.

	\setlength{\tabcolsep}{1.5mm}
\renewcommand{\arraystretch}{1.}
\begin{table}[t]
	\caption{Primary partitions in the SRE21 Audio TEST set}
	\label{tab:partitions}
	\centering{
	\begin{tabular}{llcc}
		\toprule
		\textbf{Partition} & \textbf{Elements} & \textbf{\#Target} & \textbf{\#Non-target} \\
		\midrule
		\multirow{2}{*}{Gender} & male &  25,216 & 406,731 \\
								& female & 89,785 & 4,773,077 \\
		\hline
		\multirow{2}{*}{\#enrollment} & 1 & 115,001 & 5,179,808 \\
							segments  & 3$^\dagger$ & 17,037 & 719,171 \\
		\hline
		\multirow{2}{*}{Phone\# match} & Y & 48,838 & 0 \\
									   & N & 66,163 & 5,179,808 \\
		\hline
		\multirow{2}{*}{Source type match} & Y & 51,871 & 2,322,545 \\
								  & N & 63,130 & 2,857,263 \\
		\hline
		\multirow{2}{*}{Language match} & Y & 39,509 & 2,641,177 \\
								  & N & 75,499 & 2,538,631 \\
		\bottomrule
	\end{tabular}}
\footnotesize{$^\dagger$not used in calculation of the primary metric}\\
\vspace{-2mm}
\end{table}

	In addition to the actual $C_{Primary}$, a minimum detection cost was computed by using the detection thresholds that minimize the detection cost. For minimum cost calculations, the counts for each condition set were equalized before pooling and cost calculation (i.e., minimum cost was computed using a single threshold not one per condition set).
    
    \section{Baseline Systems}
	\label{sec:baseline}
    
	\begin{figure}[b]
    \centering
    \includegraphics[scale=0.35, clip, trim=0mm 0mm 0mm 0mm]{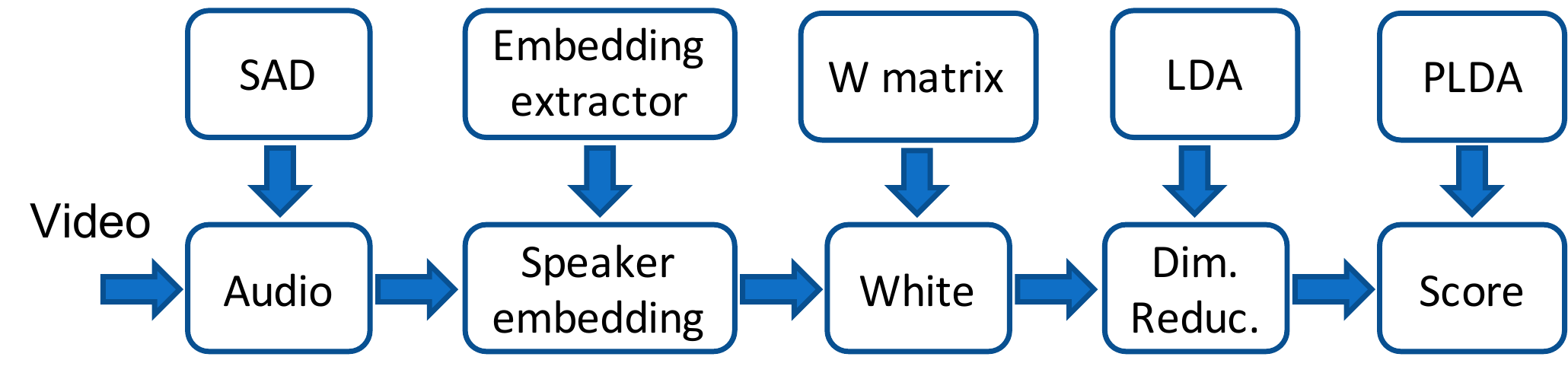}
    \caption{{\it A simplified block diagram of the SRE21 baseline speaker recognition system.}}
    \label{fig:blockdiag_xvec}
    \vspace{-2mm}
    \end{figure}

    \subsection{Speaker recognition system}

    In this section, we describe the baseline speaker recognition system setup including speech and non-speech data used for training the system components as well as the hyper-parameter configurations used in our evaluations. Figure~\ref{fig:blockdiag_xvec} shows a block diagram of the x-vector baseline system. The embedding extractor was trained using Pytorch\footnote{https://github.com/pytorch/pytorch}, while the NIST SLRE toolkit was used for front-end processing and back-end scoring.
    
    \subsubsection{Data}
    
    The baseline speaker recognition system was developed using the NIST SRE CTS Superset (LDC2021E08) \cite{supersetldc} and VoxCeleb datasets. We down-sampled all data to 8~kHz, and trained a single system for CTS and AfV trials. In order to increase the diversity of the acoustic conditions in the training set, two different data augmentation strategies were adopted. The first strategy used noise-degraded versions of the original recordings (using babble, general noise, and music), while the second strategy used spectro-temporal masking applied directly on spectrograms (aka spectrogram augmentation \cite{specaug}). The noise samples for the first augmentation approach were extracted from the MUSAN corpus \cite{musan}. For spectrogram augmentation, the mild and strong policies described in \cite{specaug} were used.
    
    \subsubsection{Configuration}
    
    For speech parameterization, we extracted 64-dimensional log-mel spectrograms from 25 ms frames every 10 ms using a 64-channel mel-scale filterbank spanning the frequency range 80~Hz--3800~Hz. After dropping the non-speech frames using SAD, a short-time cepstral mean subtraction was applied over a 3-second sliding window.
    
    For embedding extraction, an extended TDNN \cite{snyder2019} with 11 hidden layers and parametric rectified linear unit (PReLU) activation functions was trained to discriminate among the nearly 14000 speakers in the combined CTS Superset and VoxCeleb set. A cosine loss with additive margin \cite{cosface} was used in the output layer (with $m=0.2$ and $s=40$). The first 9 hidden layers operated at frame-level, while the last 2 operated at segment-level. There was a 3000-dimensional statistics pooling layer between the frame-level and segment-level layers that accumulated all frame-level outputs from the 9\textsuperscript{th} layer and computed the mean and standard deviation over all frames for an input segment. The model was trained using Pytorch and the stochastic gradient descent (SGD) optimizer with momentum ($0.9$), an initial learning rate of $10^{-1}$, and a batch size of $512$. The learning rate remained constant for the first $5$ epochs, after which it was halved every other epoch. 
    
    To train the network, a speaker-balanced sampling strategy was implemented where in each batch 512 unique speakers were selected, without replacement, from the pool of training speakers. Then, for each speaker, a random speech segment was selected from which a 400-frame (corresponding to 4 seconds) chunk was extracted for training. This process was repeated until the training samples were exhausted.
    
    After training, embeddings were extracted from the 512-dimensional affine component of the 10\textsuperscript{th} layer (i.e., the first segment-level layer). Prior to dimensionality reduction through linear discriminant analysis (LDA) to 250, 512-dimensional embeddings were centered, whitened, and unit-length normalized. The centering and whitening statistics were computed using the in-domain development data (i.e., LDC2021E09). For backend scoring, a Gaussian probabilistic LDA (PLDA) model with a full-rank Eigenvoice subspace was trained using the embeddings extracted from 600~k speech segments in the CTS Superset. The PLDA parameters were then adapted to the in-domain development data (i.e., LDC2021E09) using Bayesian maximum \textit{a posteriori} (MAP) estimation. No score normalization or calibration was applied.
    
    It is worth emphasizing that the configuration parameters employed to build the baseline system are commonly used by the SRE community, and no attempt was made to tune the hyperparameters or data lists utilized to train the models.
	
	\begin{figure}[t]
		\centering
		\includegraphics[width=0.62\linewidth, clip, trim=0mm 0mm 0mm 0mm]{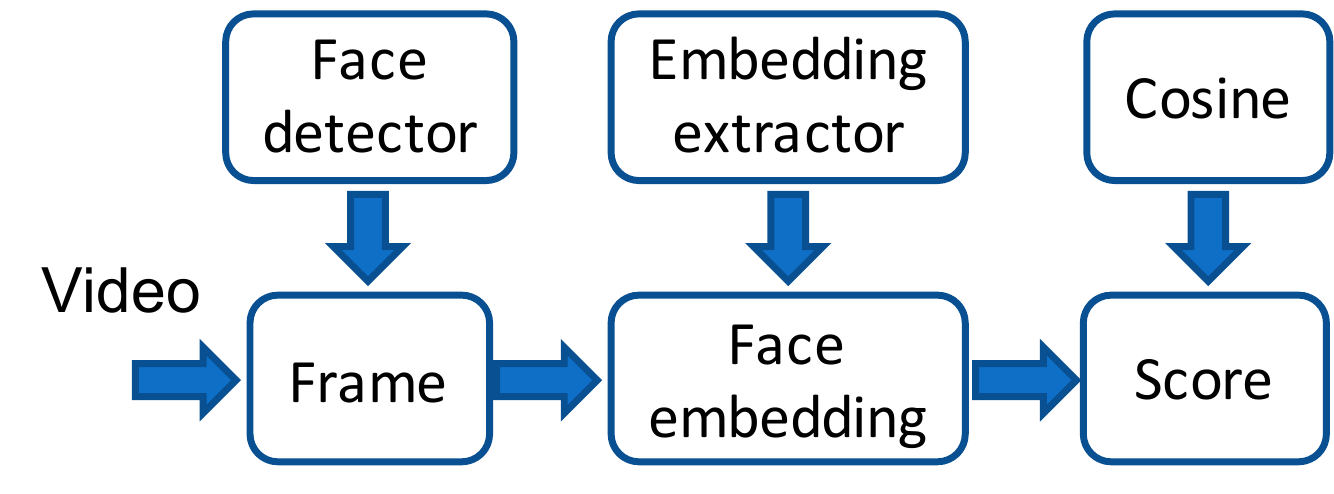}
		\vspace{-2mm}
		\caption{{\it A simplified block diagram of the SRE21 baseline face recognition system.}}
		\label{fig:blockdiag_facerec}
		\vspace{-4mm}
	\end{figure}
	
	\subsection{Face recognition}
	In this section, we describe the baseline face recognition system setup including the visual data used for training the system components as well as the hyper-parameter configurations used in our experiments. Figure~\ref{fig:blockdiag_facerec} shows a block diagram of the baseline face recognition system which was built using Pytorch based implementations\footnote{\url{https://github.com/foamliu/InsightFace-PyTorch}} of the InsightFace\footnote{\url{https://github.com/deepinsight/insightface}} with 1) a face detector termed RetinaFace \cite{Deng2020CVPR}, and 2) a face embedding extractor using a ResNet101 architecture. We used the NIST SLRE toolkit for back-end scoring.
	
	\subsubsection{Data}
    The baseline face recognition system utilized a pre-trained model available at \url{https://github.com/foamliu/InsightFace-PyTorch/releases/tag/v1.0} (model name: insight-face-v3.pt) which was trained on MS-Celeb-1M dataset \cite{guo2016ms} using a ResNet101 architecture \cite{inceptionresnet}. MS-Celeb-1M is a large-scale dataset containing approximately 3.8 million images from more than 85000 subjects spanning a wide range of different ethnicities, professions and ages.

    \subsubsection{Configuration}
    We began processing the test videos by extracting one frame per second using \texttt{ffmpeg}. Then, we applied the RetinaFace face detector on the extracted frames to 1) filter out frames with no faces, and 2) compute the bounding boxes for all faces present in each frame. Next, the face images were cropped using the bounding box coordinates, aligned using a 5-point facial landmark model, resized to $112\times112$ pixels, and normalized. Finally, the ResNet101 model was used to extract face encodings from the cropped, aligned, and normalized resized images.

    For enrollment, we used the face encodings extracted from the close-up image of each subject to build a model for that target individual. In order to compute a single score for each trial involving an enrollment image and a test video, we computed the maximum of the cosine similarity scores obtained by comparing the enrollment encoding and pseudo-test encodings. The pseudo encodings were obtained by clustering the original test encodings per video using the \textit{k-means++} algorithm \cite{kmeanspp}.
    
    \section{System Highlights}
    Before presenting the results and analysis, it is worth briefly looking at some key highlights from the top performing submissions for SRE21. In terms data usage, a majority of the leading systems used the data specified for the \textit{fixed} training condition. For the \textit{open} condition, a few teams also used in-house datasets some including recordings from more 200k speakers. Teams used extensive data augmentation to increase the diversity of the available data. In terms of core system components, all systems used neural embeddings as speaker representations; the embeddings were extracted from different variations of ResNets~\cite{he2016deep} trained using different flavors of angular margin losses (e.g., \cite{cosface, arcface}). Top performers also used domain-dependent training, long duration fine-tuning~\cite{romero2019fine}, along with adaptive score normalization~\cite{cumani2011}. For system combination, systems used both embedding level or/and score level fusion of multiple systems. For score calibration, teams used logistic and trial-based calibration. Finally, a few top performing teams explored promising directions including neural upsampling of telephony data, language-agnostic training, and wav2vec~\cite{wav2vec20} based features for speaker recognition.

	\section{Results and Discussion}
	
	In this section we present some key results and analyses for the SRE21 primary submissions, in terms of the minimum and actual costs as well as DET performance curves.
	
	\begin{figure}[b]
		\centering
		\includegraphics[width=\linewidth, clip, trim=0mm 10mm 0mm 0mm]{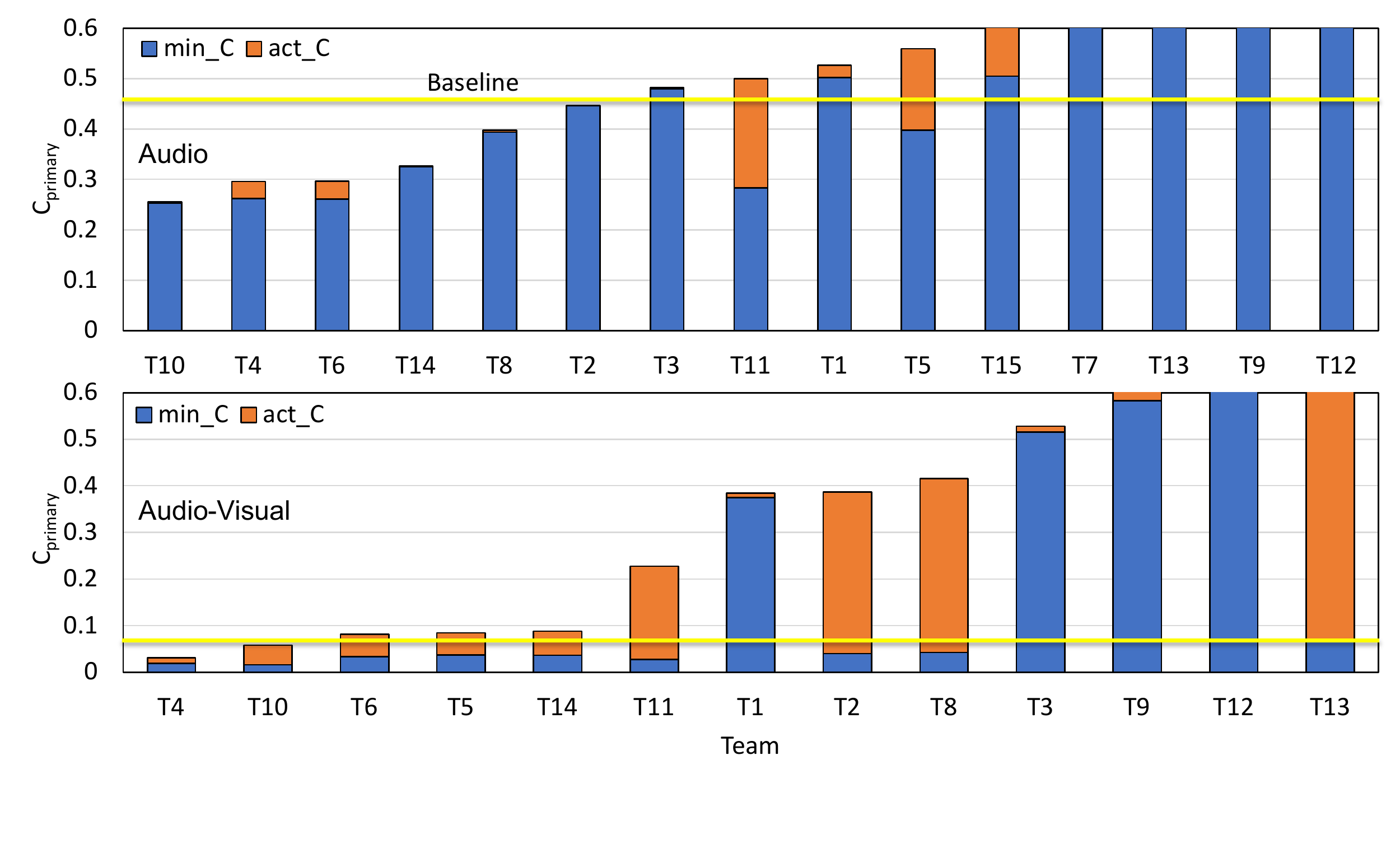}
		\vspace{-6mm}
		\caption{\it Performance of SRE21 primary submissions for the required tracks (i.e., audio and audio-visual tracks) in terms of the minimum (in blue) and actual (in orange) detection costs.}
		\label{fig:primary_results}
		\vspace{-2mm}
	\end{figure}
	
	\begin{figure}[t]
		\centering
		\includegraphics[width=\linewidth, clip, trim=0mm 0mm 0mm 0mm]{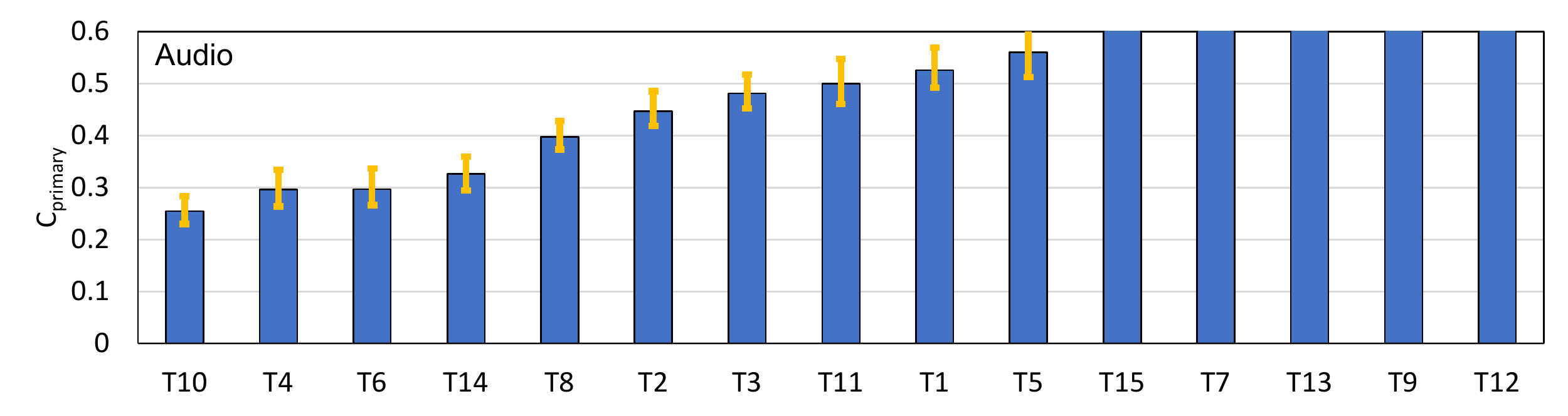}
		\includegraphics[width=\linewidth, clip, trim=0mm 5mm 0mm 0mm]{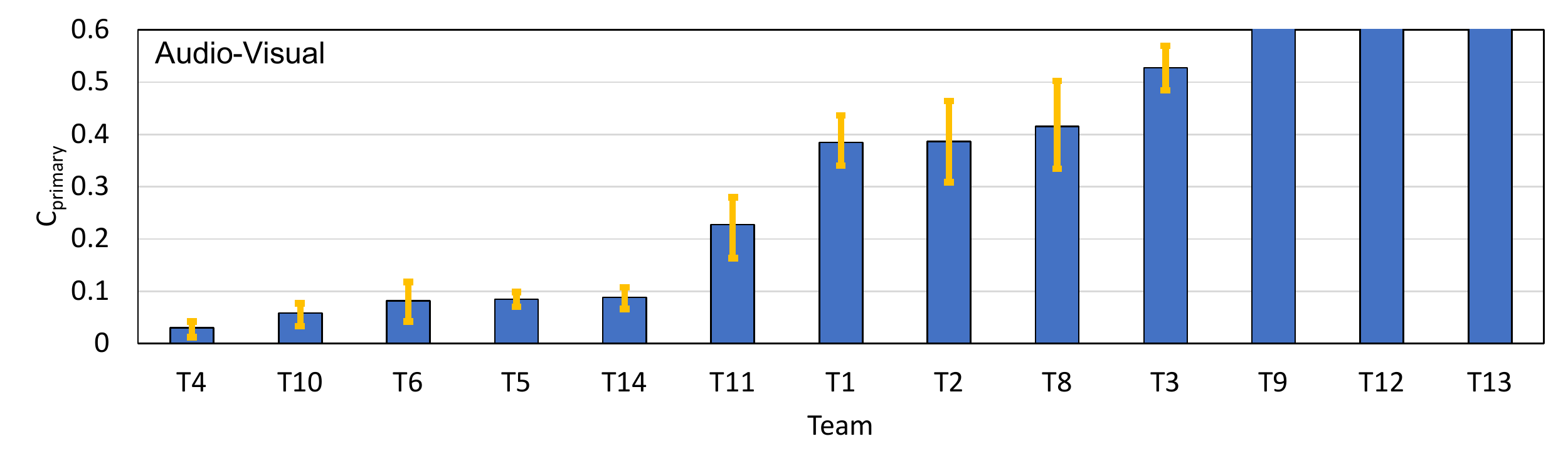}
		\vspace{-6mm}
		\caption{\it Performance confidence intervals (95\%) of the SRE21 primary submissions for the audio (top), and audio-visual (bottom) tracks.}
	\label{fig:confidence_intervals}
		\vspace{-6mm}
	\end{figure}
	
	Figure~\ref{fig:primary_results} shows the performance of the primary submissions per team for the core tracks in SRE21, i.e., audio (top) and audio-visual (bottom) tracks in terms of the actual and minimum costs for the SRE21 \textit{TEST} set. The yellow horizontal line denotes the performance of the baseline systems (see Section~\ref{sec:baseline}) in terms of the minimum cost. The y-axis limit is set to $0.6$ to facilitate cross-system comparisons in the lower cost region. The blue bars represent the minimum of the decision cost function, and the orange bars show the actual costs computed for each system at the threshold values defined in Section~\ref{sec:metric}. Several observations can be made from this figure. First, the top performing teams outperform the baseline, with the top performers achieving more than 44\% improvement over the baseline. Second, for the audio track, a majority of top performers achieved good calibration performance given the challenges introduced due to cross-lingual and cross-source trials. Third, for the audio-visual track, we see significant reduction in $C_{primary}$ due to fusion, but the calibration errors remain high. Based on our analysis, these calibration errors are primarily caused by the miscalibration issue for the visual systems. Similar to the audio track, a majority of the systems outperformed the audio-visual baseline system in terms of minimum $C_{primary}$. Notice that the order of the top performing systems in the audio track is not necessarily the same as that of the audio-visual track.
	
	
	It is common practice in the machine learning community to perform statistical significance tests to facilitate a more meaningful cross-system performance comparison. Accordingly, similar to our SRE19 analysis and to encourage the speaker recognition community to consider significance testing while comparing systems or performing model selection, we computed bootstrapping-based 95\% confidence intervals using the approach described in \cite{bootstrap}. To achieve this, we sampled, with repetition, the unique speaker model space along with the associated test segments 1,000 times, which resulted in 1,000 actual detection costs, based on which we calculated the quantiles corresponding to the 95\% confidence margin. Figure~\ref{fig:confidence_intervals} shows the performance confidence intervals (around the actual detection costs) for each team for the audio (top), and audio-visual (bottom) tracks. It is seen that the top performing system in the audio track is significantly better than the rest of the systems, and that the next 3 top performers are not significantly different than each other. For the audio-visual track, we observe that not only does the multimodal fusion improve the actual $C_{primary}$, but also it shrinks the confidence intervals for the top performing systems. It is also interesting to observe that for some systems the confidence intervals are much narrower compared to the others which could indicate robustness to the various samplings of the trial space. These observations further highlight the importance of statistical significance tests while reporting performance results or in the model selection stage during system development, in particular when the number of trials is relatively small.
	
	\begin{figure}[t]
		\centering
		\includegraphics[width=.95\linewidth, clip, trim=0mm 5mm 0mm 0mm]{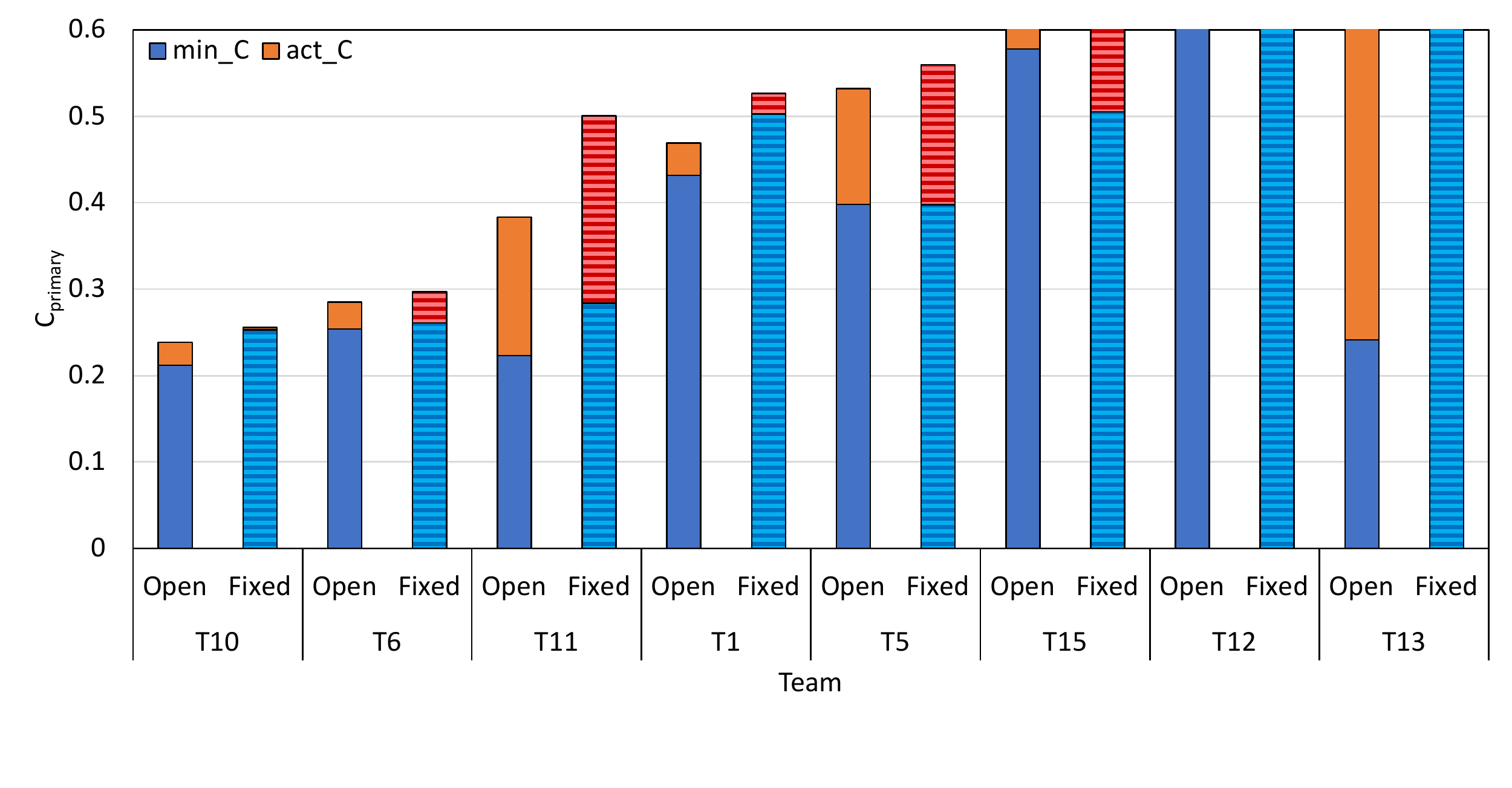}
		\vspace{-4mm}
		\caption{\it Performance of the SRE21 primary \textit{fixed} vs \textit{open} submissions for the audio track.}
	\label{fig:open_vs_fixed}
		\vspace{-4mm}
	\end{figure}
	
		\begin{figure}[t]
		\centering
		\includegraphics[width=.8\linewidth, clip, trim=0mm 0mm 0mm 15mm]{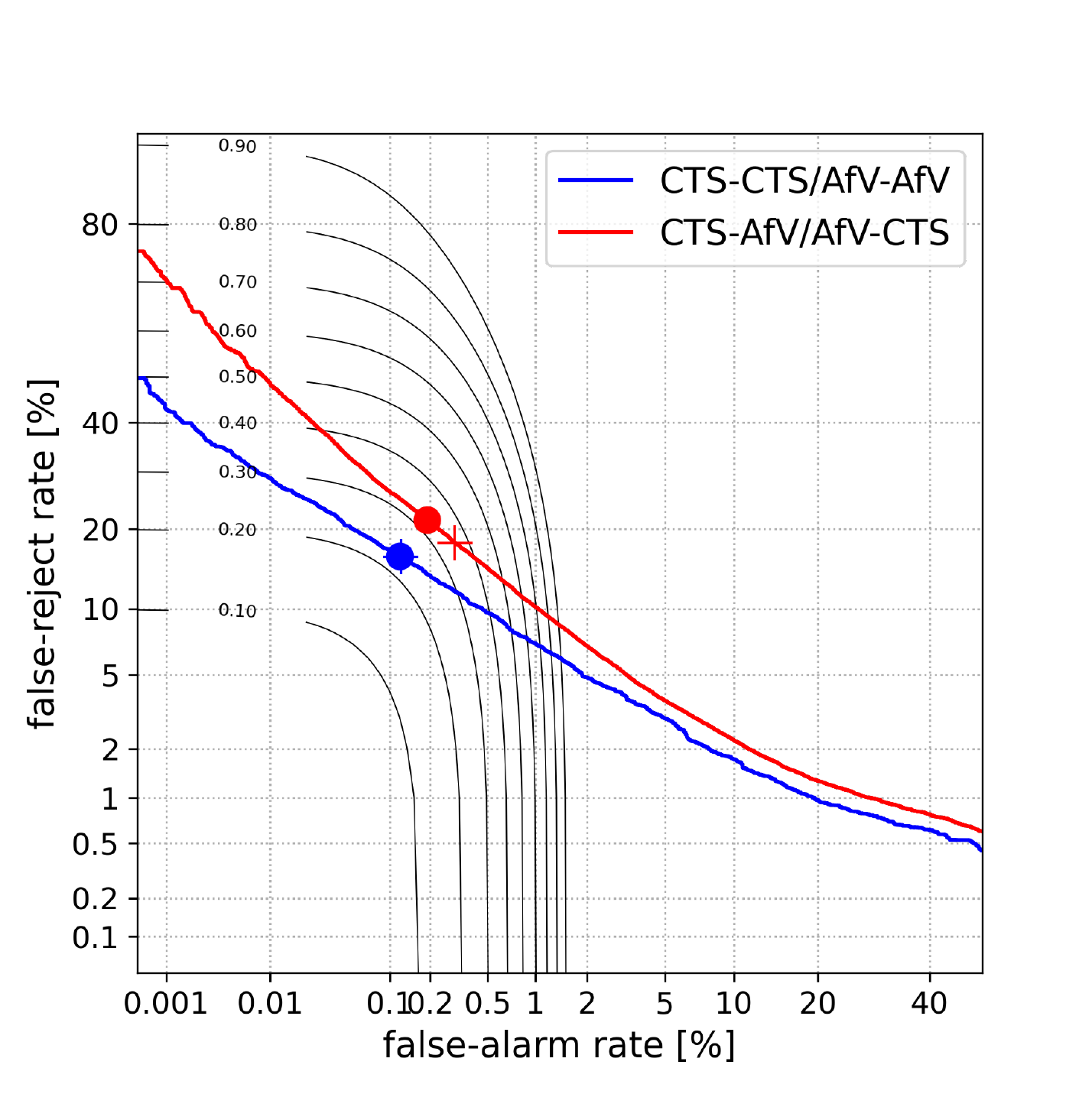}
		\vspace{-2mm}
		\caption{\it DET curve performance of a top performing system by source type match, for the \textbf{audio} track.}
	\label{fig:source_type_match}
		\vspace{-4mm}
	\end{figure}
	
	\begin{figure}[t]
		\centering
		\includegraphics[width=.8\linewidth, clip, trim=0mm 0mm 0mm 15mm]{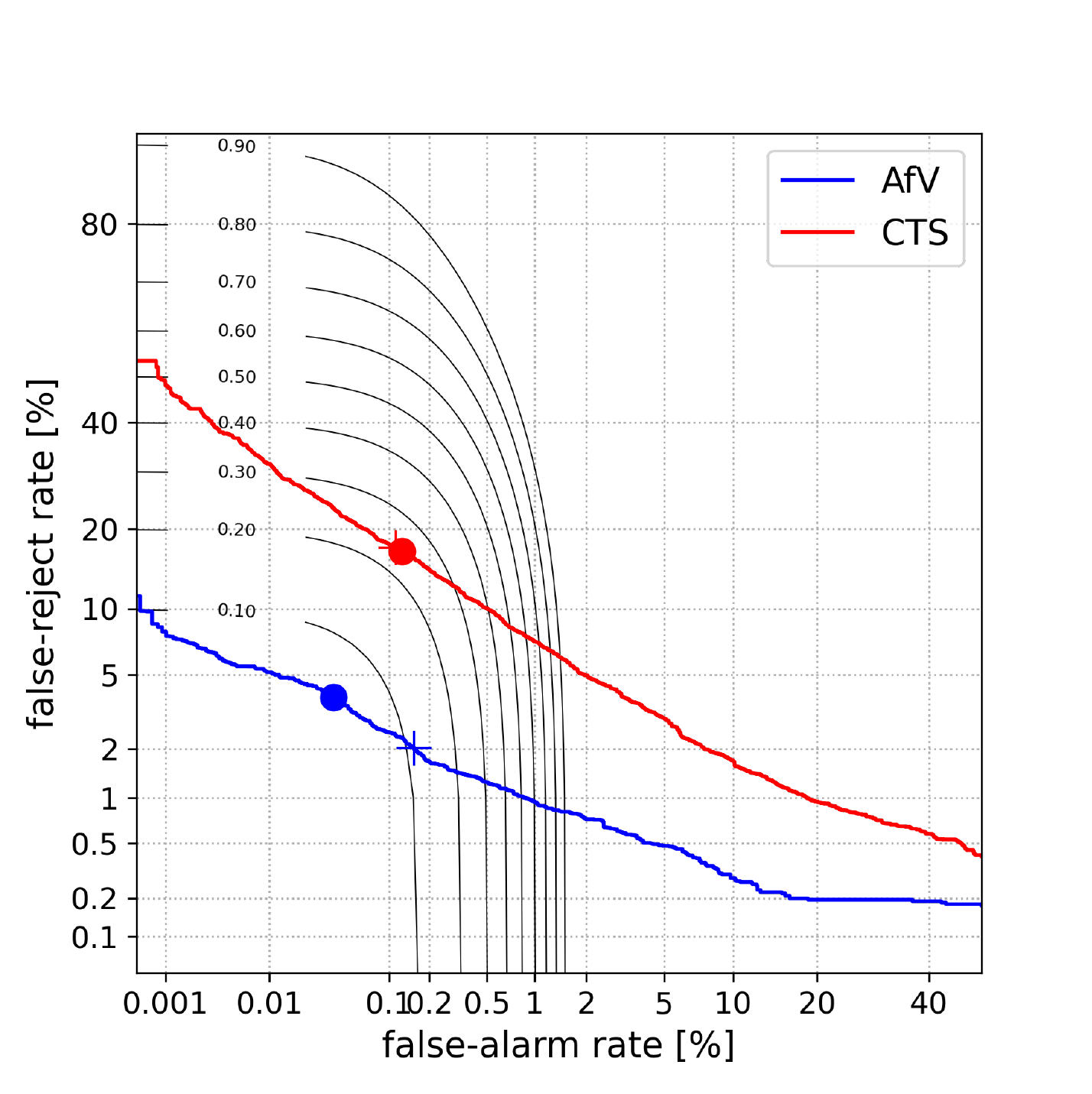}
		\vspace{-3mm}
		\caption{\it DET curve performance of a top performing system by source type (i.e., CTS vs AfV), for the matched source condition in the \textbf{audio} track.}
	\label{fig:cts_vs_afv}
		\vspace{-4mm}
	\end{figure}
	
	Figure~\ref{fig:open_vs_fixed} shows a breakdown of performance by training condition for the audio track. The \textit{open} training condition seems to result in slight improvements in performance, in particular for T11 and T1 systems.

	Figure~\ref{fig:source_type_match} shows DET performance curves of a top performing system by source type match in enrollment versus test, for the audio track. The closer the curves to the origin, the better the performance. The circles and crosses denote the minimum and actual costs, respectively, while the solid black curves represent the equi-cost contours, meaning that all points on each curve correspond to the same cost value. As expected, we observe better performance when there is a source type match between enrollment and test across all operating points. However, it is interesting to note that the gap in performance remains relatively small in the high false-alarm region.
	
	Focusing on the matched source type, Figure~\ref{fig:cts_vs_afv} shows the performance differences for the two source types, CTS vs AfV, for a top performing system. We observe substantially better performance  for AfV vs CTS. This is attributed to 1) larger bandwidth and quality for the AfV vs CTS (that is 16 kHz vs 8 kHz), and 2) relatively static conditions in the videos in terms of both background and speaker variabilities.
	
	\begin{figure}[t]
		\centering
		\includegraphics[width=.8\linewidth, clip, trim=0mm 0mm 0mm 15mm]{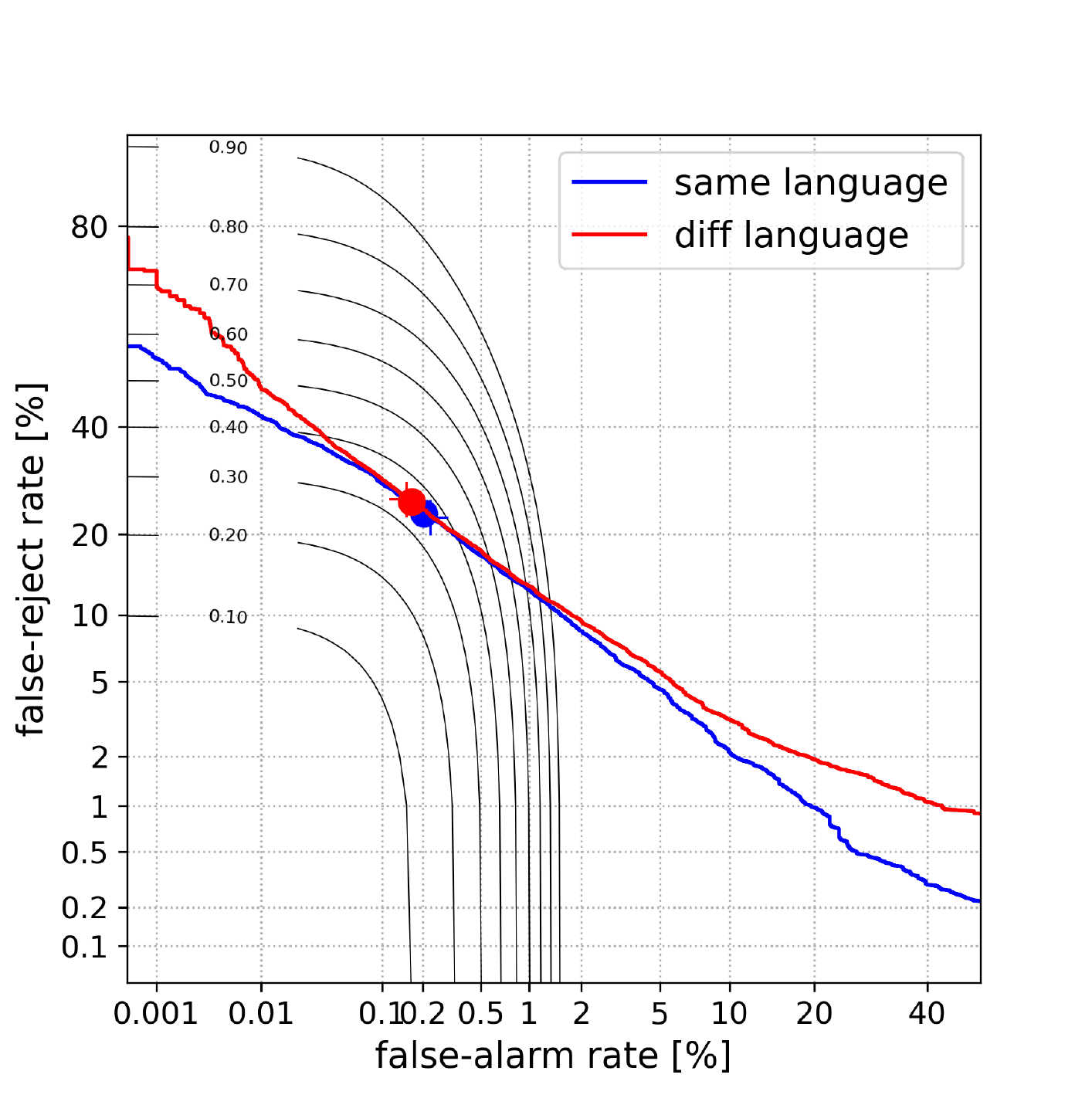}
		\vspace{-3mm}
		\caption{\it DET curve performance of a top performing system by language match, for the \textbf{audio} track.}
	\label{fig:language_match}
		\vspace{-5mm}
	\end{figure}
	
	Figure~\ref{fig:language_match} shows the DET performance plot of a top performing system by language match for the audio track. In general performance is better when the same language is spoken in the enrollment and test segments; however, the gap is relatively small in particular in the vicinity of the operating points of interest (i.e., the low false-alarm region).
	
	\begin{figure}[t]
		\centering
		\includegraphics[width=.8\linewidth, clip, trim=0mm 0mm 0mm 15mm]{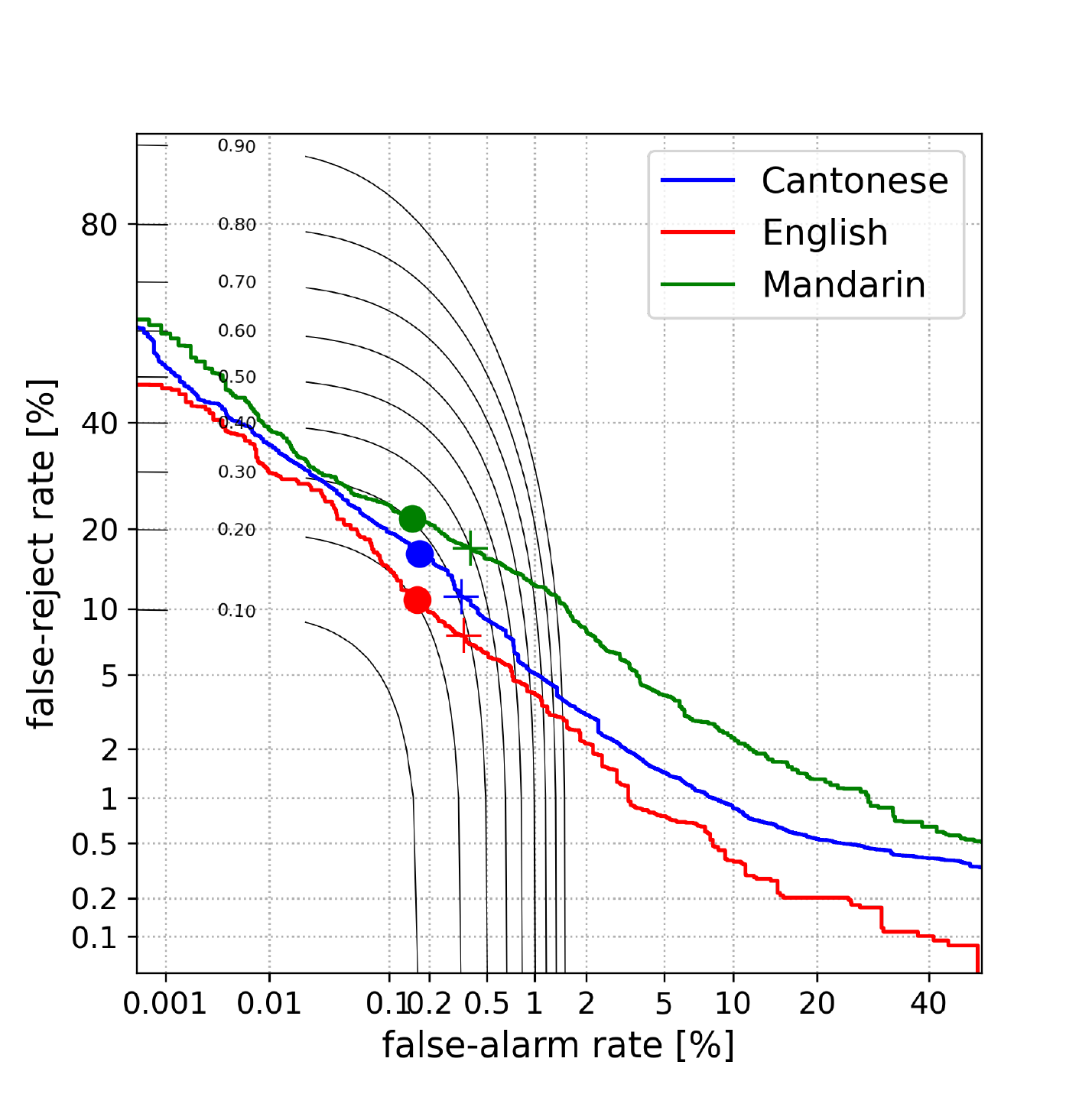}
		\vspace{-3mm}
		\caption{\it DET curve performance of a top performing system by language, for the \textbf{audio} track.}
	\label{fig:language}
		\vspace{-4mm}
	\end{figure}
	
	We also analyzed the performance by the 3 languages in the SRE21 audio \textit{TEST} data. Figure~\ref{fig:language} shows DET performance curves of a top performing system for Cantonese, English, and Mandarin. Overall, we observe better performance for English, followed by Cantonese and Mandarin. The better performance on English is justified by the availability of large amounts of English data for system training in SRE’21. Nevertheless, it remains unclear why Mandarin is the most challenging language.
	
	\begin{figure}[t]
		\centering
		\includegraphics[width=.8\linewidth, clip, trim=0mm 0mm 0mm 7mm]{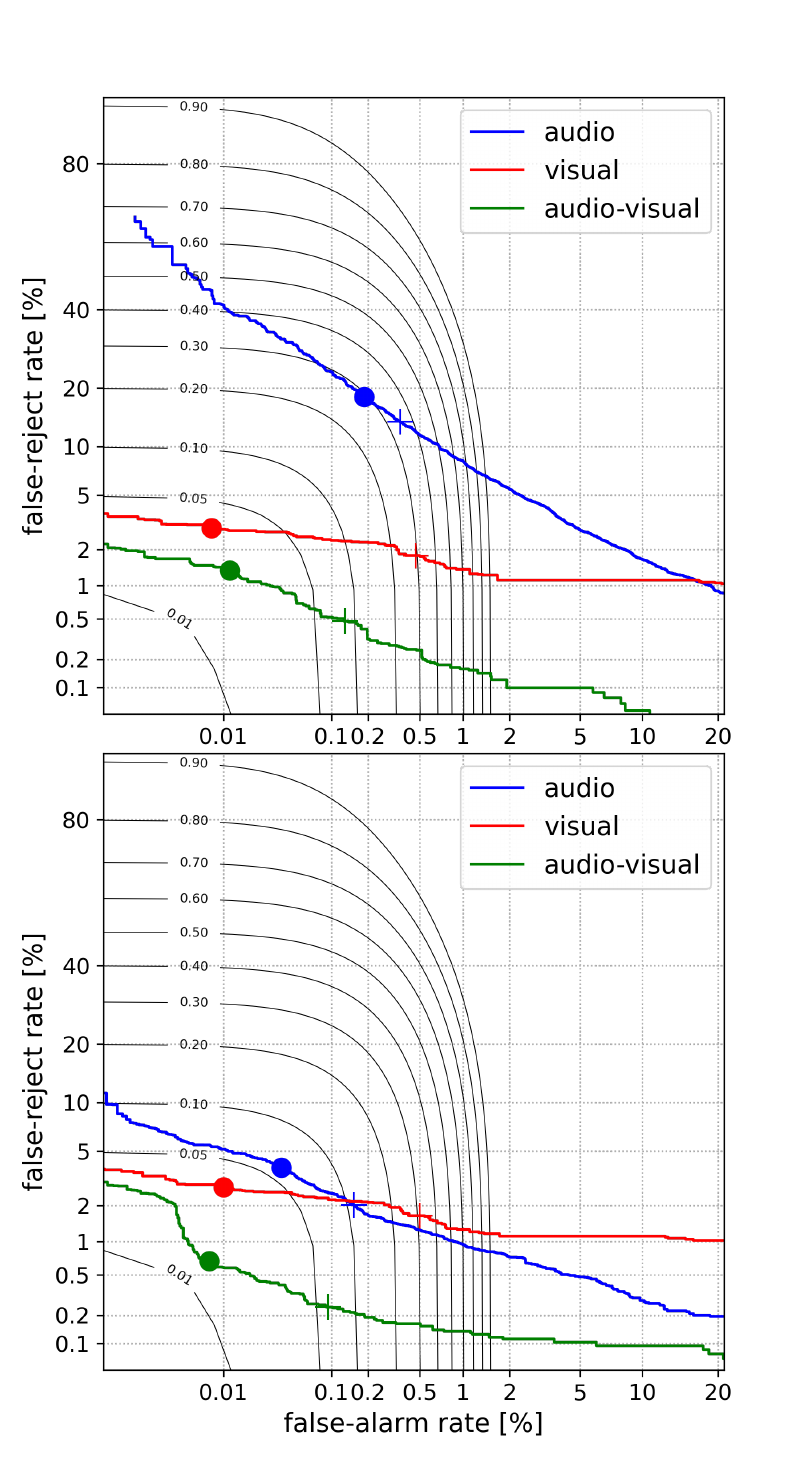}
		\vspace{-3mm}
		\caption{\it DET curve performance of the top performing system for the \textbf{audio}, \textbf{visual}, and \textbf{audio-visual} tracks, for mismatched (top) and matched (bottom) enrollment-test conditions.}
	\label{fig:fusion}
	\end{figure}
	
	Figure~\ref{fig:fusion} shows DET performance plots by the two modalities (i.e., audio and visual) as well as their fusion for a top performing system. The top plot shows the DET curves for the mismatched enrollment-test scenario (CTS vs AfV), while the bottom plot shows the curves for the matched scenario (AfV vs AfV). We observe that even when there are large gaps in performance for one modality versus the other, fusion results in large gains in performance, which is expected given the complementarity of the two modalities. In the bottom plot, we still observe large gains from fusion, but we also see that the top performing audio and visual systems perform comparably in the operating region of interest which is a promising sign for the speaker recognition community.
	
	\section{Conclusion}
	SRE21 was the second large-scale audio-visual speaker recognition evaluation conducted by NIST. Thanks to a new multimodal and multilingual dataset collected by the LDC, termed WeCanTalk, SRE21 introduced new challenges for the research community including cross-lingual and cross-domain target and non-target trials. In the cross-lingual scenario, the enrollment and test segments were spoken in different languages, while in the cross-domain (or cross-source) scenario the enrollment and test segments originated from different source types (i.e., CTS and AfV). In this paper, we presented an overview of SRE21 including the task, data, performance metric, baseline systems, as well as results and system performance analyses. The main observations are as follows: 1) audio-visual fusion results in substantial gains in speaker/person recognition performance, 2) top performing matched domain speaker (AfV-AfV) and face recognition systems seem to perform comparably for the person recognition task\footnote{We stress here that the results utilizing face recognition are part of the research described in this paper, conducted while developing the NIST SRE's, and these results are not part of the NIST Facial Recognition Vendor Test (FRVT).}, and 3) speaker recognition performance improvements were largely attributed to the use of ResNets with angular margin losss, data augmentation, and condition-dependent system development.
	
	\section{Disclaimer}
	
	These results presented in this paper are not to be construed or represented as endorsements of any participant's system, methods, or commercial product, or as official findings on the part of NIST or the U.S. Government.
	
	\section{Acknowledgement}
	
	Experiments and analyses were performed, in part, on the NIST Enki HPC cluser.
	
	\balance
	\bibliographystyle{IEEEtran}
	\bibliography{Odyssey2020_BibEntries}
	
\end{document}